\documentclass[12pt]{article}           
\def\preprint{}       
\def\finished{} %January 2000}

\def\title{Redshift and gauge choice}

\long\def\abstract{
  We show that a specific gauge choice comes extremely close to defining
  a frame whose preferred observers see a dipole-free CMB.
  In this gauge the metric is the product of a scale factor
  depending on all spacetime coordinates, and a metric featuring an
  expansion-free geodesic timelike vector field.
  This setup facilitates the computation of redshift and other distance
  measures and explains why we can have a highly isotropic CMB
  despite large inhomogeneities.
}

\def\gh{{\hat g}}   \def\uh{{\hat u}}
 \def\sch{{\hat ;}}  \def\Gh{{\hat \Gamma}}
  \def\ah{a_\mathrm{h}}

\def\ddf{\delta_\mathrm{df}}  \def\dlg{\delta_\mathrm{lg}}
\def\udf{u_\mathrm{df}} 

\input epsf

\usepackage{amssymb,latexsym}              \catcode`\"=\active \let"=\"

\def\ifundefined#1{\expandafter\ifx\csname#1\endcsname\relax}
\def\bye{\end{document}}   
\long\def\new#1\endnew{{\bf #1}}
\long\def\del#1\enddel{} 

\def\HS#1 {\hspace*{#1pt}} \def\VS#1 {\vspace*{#1pt}}
\def\BC{\begin{center}}    
\def\EC{\end{center}}

\let\0=\over      \let\1=\vec      \def\2{{1\over2}}    \let\3=\ss
\let\4=\underline \let\5=\overline \let\6=\partial      \def\7#1{{#1}\llap{/}}
\def\8#1{{\textstyle{#1}}}         \def\9#1{{\ifmmode{\pmb{#1}}\else\bf#1\fi}}

          \def\({\left(}       \def\){\right)}

\def\eeql#1 {\label{#1}\eeq}        
\def\beq{\begin{equation}}      \def\eeq{\end{equation}}        
\def\bea{\begin{eqnarray}}      \def\eea{\end{eqnarray}} 

\def\mao#1{\mathop{\rm #1}\nolimits}  
\def\tr{\mao{tr}}

\let\and=\wedge

\let\bra=\langle        \let\ket=\rangle        \def\<#1\>{\bra #1 \ket}

\def\rel#1 #2{\buildrel #1 \over {#2}}  
\def\fnote#1#2{\begingroup\def\thefootnote{#1}\footnote{#2}
                \addtocounter{footnote}{-1}\endgroup}   

       \let\g=\gamma   \let\d=\delta   %% --> GREEK
           
   \let\l=\lambda  \let\m=\mu      
\let\n=\nu                  \let\r=\rho     \let\s=\sigma 
     \let\o=\omega            
            \let\O=\Omega    
        \let\L=\Lambda     \let\D=\Delta

\ifundefined{draftmode} {} 
\else        
   \newcount\HOUR  \HOUR=\the\time \divide\HOUR by 60   \multiply \HOUR by 60 
   \newcount\MIN   \MIN=\the\time  \advance\MIN by -\HOUR  \divide\HOUR by 60
   \ifnum  \MIN>9  \def\printTIME{{\it\the\HOUR\,:\,\the\MIN}}
   \else   \def\printTIME{{\it\the\HOUR\,:\,0\the\MIN}} 
   \fi % \printTIME
   
   \def\LLab#1{\BP(0,0)\unitlength=1mm\put(-12,.5){\makebox(0,0)[cr]{\small #1
        \rlap{$_{_{\makeatletter\csname TRef#1\endcsname\makeatother}}$}}}\EP}
\fi

\begin{document}

%\del

{\hfill\preprint }
\vskip 15mm
\begin{center} 
{\huge\bf   \title }\vskip 10mm
Harald Skarke\fnote{*}{e-mail: skarke@hep.itp.tuwien.ac.at}\\[3mm]
Institut f\"ur Theoretische Physik, Technische Universit\"at Wien\\
Wiedner Hauptstra\ss e 8--10, 1040 Wien, Austria
        
\vfill                  {\bf ABSTRACT } 
\end{center}    
\abstract

\vfill \noindent \preprint\\[5pt] \finished \vspace*{9mm}
\thispagestyle{empty} \newpage
\pagestyle{plain}

\newpage
\setcounter{page}{1}
%\enddel

\section{Introduction}

The almost perfect isotropy of the cosmic microwave background (CMB)
is among the pillars of the cosmological standard model %which states that
according to which
our universe can be described, at large scales, as a 
Friedmann-Lemaitre-Robertson-Walker (FLRW) universe with small perturbations.
This isotropy comes at different levels (see \cite{Akrami:2018vks} for CMB
data from Planck and \cite{Hoffman:2015waa} for the peculiar velocities, or
\cite{Tanabashi:2018oca} for a useful summary): %pdg
the actual observations (terrestrial or from satellites)
%moving non-relativistically)
show deviations in temperature of $\d T/T \approx 0.12 \%$, % \cite{bla},
but once the dipole contribution is subtracted,
this improves to a value of $\ddf \approx 10^{-5}$ (here and in the following,
we abbreviate $\d T/T$ by $\d$ and use subscripts such as `df' for
`dipole-free' to indicate which observer we are referring to).
%\cite{bla}.
%In other words, while we see homogeneity at the level of $2\times 10^{-2}$,
This means that an observer passing through our solar system at a velocity
of $370$\,km/sec (in the right direction) will see the latter spectacularly
small level;
on the other hand, an observer comoving with our local galaxy group sees
an anisotropy of $\dlg\approx 0.2\%$.

According to the Copernican principle, the situation should be similar at most
locations in the present era.
It is important to note the difference between $\ddf$ and $\dlg$,
not only in size
($\ddf \approx 10^{-5}\ll \dlg \approx 2\times 10^{-3}$), but also in quality:
whereas $\ddf$ is determined by a full celestial sphere's worth of observations,
%of $\d\l/\bar \l$,
the value of $\dlg$ comes from a single draw from a distribution with mean
zero.
For these reasons, we would very much prefer the use of $\ddf$ over that of
$\dlg$ in an analysis of the structure of the universe.
In other words, we want to work in a frame comoving with the CMB, not with the
matter.

%Any fluctuation in the CMB temperature comes from a combination of
The wavelength of a CMB photon is the product of its value at last 
%original temperature
scattering and the redshift factor %that %every photon experiences
picked up on the way to the observer.
Unless one believes in strange nonlocal correlations between the two,
one can only conclude that neither the original wavelength nor the redshift
factor should feature deviations that are larger than the ones seen by the
observer.
In the present work we will be interested only in the extremely precise
matching of the redshifts in the different directions.

The celebrated Ehlers-Geren-Sachs (EGS) theorem \cite{Ehlers:1966ad} states
that the existence of a perfectly isotropic radiation background combined with
reasonable assumptions on the matter content of the universe implies FLRW.
There is a number of generalizations to `almost EGS' theorems (e.g.\
\cite{Stoeger:1994qs,Clarkson:1999yj,Rasanen:2009mg,Clarkson:2010uz})
stating that small deviations from isotropy should lead only to small deviations
from FLRW; see section 11.1 of Ref.\ \cite{emm} for a very clear summary.
These works usually (with an exception in \cite{Rasanen:2009mg})
assume that the radiation 4-velocity (i.e.\ the
velocity field $\udf$ of the dipole-free observers) is geodesic.
This is an additional input which can be argued for only if one %identifies
does not distinguish the CMB frame from the matter frame.
Thus it holds only at the level of $\dlg$, not at the level of $\ddf$.

In the present work we are interested in precision at the level of
$\ddf \approx 10^{-5}$, so we do \emph{not} take the radiation velocity
to be geodesic.
Our analysis will rely on redshift rather than distribution functions for
the radiation, which simplifies matters considerably.
The %condition of a vanishing CMB dipole defines a
timelike vector field $\udf$ that determines a preferred observer
at every spacetime point can, in principle, be completed to an
orthonormal frame $\{e_0 = \udf,\,e_1,\,e_2,\,e_3\}$ which we would call %such a frame
a CMB frame.
In practice the requirement of a vanishing dipole is highly nonlocal and
%makes such a frame
therefore analytically intractable.
Instead, we are going to work with a locally well-defined quantity which,
as we shall explicitly verify, comes very close to %giving us a CMB frame.
defining the level of anisotropy.
It turns out that this quantity can be simplified by a conformal
transformation, and that the most important contributions to it can be
eliminated by a gauge choice.
%The possibility of eliminating
%This explains why
The physical observable $\ddf$ is of course gauge invariant and can therefore
be computed in any gauge. 
Choosing the one suggested here makes it particularly transparent why
$\ddf$ is so small despite the fact that the actual
universe shows a considerable amount of inhomogeneity.
Working in this gauge significantly improves the tractability of light
propagation compared to the synchronous and the longitudinal gauge, which
are the ones that are used most frequently.
An explicit comparison in linear perturbation theory shows that the metric
perturbations in the new gauge are not much larger than those in the
longitudinal gauge, which is usually considered to be optimal in that respect.

In the next section we introduce a
%condition that we are going to use as a local proxy for defining
quantity that vanishes %for a homogeneously
if an isotropically redshifted CMB is observed everywhere, and show how it
simplifies under a conformal transformation.
In section 3 we formulate a gauge that eliminates two  of three
contributions to this quantity and thereby comes close to defining a
CMB frame; we also give explicit conditions on a metric implementing this
gauge.
Section 4 contains an analysis of this metric in linear perturbation theory
and comparisons with other gauges.
In the final section we argue that other distance measures are also well
behaved in the new gauge, make some remarks on the controversy about the
impact of inhomogeneities on the expansion of the universe, and discuss
open questions about our gauge.

\section{Redshift and conformal transformation} % and gauge choice}
%in the absence of homogeneity}

%We denote our spacetime manifold by $\cm$ and the pseudo-Riemannian metric
%on it by $g_{\m\n}$.
We consider a photon emitted at some point $x_e$ by a source moving
along a worldline with a tangent vector $u_e$ normalized to
$u_e^2 = g_{\m\n}u_e^\m u_e^\n = -1$, where $g_{\m\n}$ is the pseudo-Riemannian
spacetime metric of type $-+++$.
This photon propagates along a lightlike geodesic which we describe by an affine
parameter $\l$ such that the tangent vector to the geodesic is
$k^\m = dx^\m / d\l $.
%\beq k^\m = \frac{dx^\m}{d\l }. \eeql{lightlike}
The redshift $z_{e\to o}$, as seen by an obvserver at $x_o$ whose wordline has
the tangent vector $u_o$ (normalized to $u_o^2 = -1$), is determined by the
well-known formula
\beq 1+z_{e\to o} = {(u\cdot k)_e\0 (u\cdot k)_o}.\eeql{redsh}

In an idealized universe in which every spacetime point admits a
distinguished observer
who sees a perfectly isotropically redshifted last scattering surface,
there would exist a global vector field $u$ characterizing such observers,
as well as a globally well defined function
\beq a(x) = 1 + z_{\mathrm{lss}\to x} = \frac{(u\cdot k)_\mathrm{lss}}{(u\cdot k)_x}
\eeql{ax}
that determines this redshift.
%In terms of this function we can
%Assuming $(u\cdot k)_\mathrm{lss}$ to be normalized to the same value on every
%point of the last scattering surface,
We could then determine the redshifts between preferred observers via
\beq 1+z_{e\to o} = \frac{a(x_o)}{a(x_e)}\eeql{redsha}
as a direct consequence of Eqs.~(\ref{redsh}) and (\ref{ax}).
Along any geodesic described with an affine parameter $\l$ and tangent vector
$k$, the value of $a(x) (u\cdot k)(x)$ would remain constant and therefore
the quantity
%Assuming a normalization of $(u\ cdot k)|_x = 1$ this implies that 
\beq d(x,k) = \frac{d}{d\l }[a(x) (u\cdot k)(x)] \eeql{dkdef}
%in the last step
would have to vanish at every spacetime point $x$ for every lightlike
tangent vector $k$ at $x$.

For an arbitrary timelike vector field $u$ and non-vanishing scalar $a$,
%with $d(x,k)$ defined as above
%A redshift formula for the case
where $d(x,k)$ need not vanish, a redshift formula can still be obtained
%in the following manner. By virtue of Eq.\ (\ref{dkdef}),
by noting that 
\beq \ln[- a(x) (u\cdot k)(x)]_e^o
  = \int_e^o \frac{d(x,k)}{a(x) (u\cdot k)(x)} d\l
  %= \int_e^o \frac{\D_{\m\n}\,k^\m\, k^\n}{a\, u_\r\,k^\r} d\l
  \eeq
%  &=& \int_e^o \frac{\hat\D_{\m\n}\,\hat k^\m\, \hat k^\n}{\hat u_\r\,\hat k^\r%}
%      d\hat\l \\
%  &=& -\int_e^o \hat \D_{\m\n}^\mathrm{T}\,\hat e^\m \, \hat e^\n d\hat\l \, ;\eea
  %and therefore
implies
\beq 1+z_{e\to o} = {(u\cdot k)_e\0 (u\cdot k)_o} =  \frac{a(x_o)}{a(x_e)}
\exp\(-\int_e^o \frac{d(x,k)}{a(x) (u_\r k^\r)(x)} d\l \)\, .
\eeql{redshfD}
In the following %we want to analyse how well the condition $d(x,k)=0$ can
%be approximated in a more general universe.
we would like to treat the requirement
\beq \< d(x,k) \> = 0, \quad \< d(x,k)^2 \> \;\mathrm{small},   \eeql{dkcond}
where $\< ~\cdots ~ \>$ should represent the average over the celestial sphere,
\beq \< ~\cdots ~\> = \frac{1}{4\pi} \int \cdots ~d\O , \eeql{celav}
as a local proxy for the conditions defining the CMB frame.
Using the facts that differentiation by $\l$ corresponds to covariant
differentiation along $k$ and that $k^\n k_{\m;\n} = 0$ we get
\beq d(x,k) = k^\n [a(x) (u\cdot k)(x)]_{,\n}
= a_{,\n} k^\n (u\cdot k) + au_{\m;\n} k^\n k^\m .
\eeql{ddlauk}
Motivated by the FLRW case, we introduce the conformally transformed quantities
\beq \gh_{\m\n} = a^{-2} g_{\m\n}, ~~~\hat u_\m = a^{-1}u_\m  ,
~~~\hat u^\m = \gh^{\m\n} \uh_\n = a u^\m  \eeql{hatted}
with $\hat u_\m \hat u_\n \gh^{\m\n} = u_\m u_\n g^{\m\n} = -1$.
Then a short calculation gives
\beq
a^2 \uh_{\m \,\hat ;\,\n} = a u_{\m;\n} + a_{,\m}u_\n - a_{,\r} u^\r g_{\m\n} ,
\eeql{cdu}
where $\hat ;$ denotes covariant differentiation with respect to $\hat g$.
%Upon normalizing $k$ to $(u\cdot k)_x = -1$ we obtain
Contraction with $k^\m k^\n$ shows that 
\beq d(x,k) = \D_{\m\n}(x)k^\m k^\n = \hat\D_{\m\n}(x)\hat k^\m \hat k^\n\eeql{dk}
with
\beq \D_{\m\n} = %(\ln a)_{,(\n} u_{\m)} + u_{(\m;\n)}
a u_{(\m;\n)} + a_{,(\m} u_{\n)} - a_{,\r} u^\r g_{\m\n} = a^2 \hat\D_{\m\n},
\quad \hat\D_{\m\n} = \uh_{(\m \,\hat ;\,\n)}.\eeql{Dmn}
%We note that the simultaneous vanishing
Thus Killing's equation $\uh_{(\m\sch\n)}=0$ implies $d(x,k)=0$,
%of all components is equivalent to, which is just.
and with a little work the converse can also be shown.
This corresponds to %we have reproduced
the well-known result \cite{Tauber:1961lbq} that a spacetime
admits a perfectly isotropic CMB background if and only if its metric
is conformal to a metric with a timelike Killing vector; this fact is
essential for the derivation of the EGS theorem \cite{Ehlers:1966ad}.

The standard decomposition (see e.g.\ chapter 4 of \cite{emm}) of
\beq g_{\m\n} = -u_\m u_\n + h_{\m\n} \eeq
into projection operators $-u_\m u_\n$ (timelike) and $h_{\m\n}$ (spacelike),
with $u^\m h_{\m\n}=0$ and $h^{\m\n}h_{\m\n}=3$,
(or, equivalently, $ \gh_{\m\n} = -\uh_\m \uh_\n + \hat h_{\m\n}$ etc.)
affords a decomposition  of any symmetric tensor $\D_{\m\n}$ as
\beq \D_{\m\n} = u_\m u_\n \D^\mathrm{St} + h_{\m\n} \D^\mathrm{Ss}
   - u_\m \D^\mathrm{V}_\n - u_\n \D^\mathrm{V}_\m + \D^\mathrm{T}_{\m\n}
\eeql{Ddecomp}
in terms of scalars ${\D^\mathrm{St}}$ and $\D^\mathrm{Ss}$ (related to the time
and space projections, respectively), a vector $\D^\mathrm{V}_\m$ satisfying
$\D^\mathrm{V}_\m u^\m=0$ and a symmetric tensor $\D^\mathrm{T}_{\m\n}$ satisfying
$\D^\mathrm{T}_{\m\n} u^\m = 0$ and $\D^\mathrm{T}_{\m\n} h^{\m\n} = 0$.

Assuming that we have parametrized the geodesic in such a way that
$u\cdot k = -1$ at the point $x$ where we compute $d(x,k)$, writing
\beq k^\m = u^\m + e^\m, \eeq
and using the conditions $u^2 = -1$ and $k^2 = 0$, %imply
we find that 
\beq u \cdot e = 0, \quad e^2 = 1\quad \hbox{and}\quad e^\m h_{\m\n} = e_\n, \eeq
i.e.\ $e$ must be a spacelike unit vector orthogonal to $u$.
Applying this to Eq.~(\ref{dk}) with the decomposition (\ref{Ddecomp}), we find
\beq d(x,k) = \D^\mathrm{S} + 2 \D^\mathrm{V}_\n e^\n + \D^\mathrm{T}_{\m\n}e^\m e^\n
\quad \hbox{with} \quad \D^\mathrm{S} = \D^\mathrm{St} + \D^\mathrm{Ss}.  \eeql{dDe}

In order to evaluate averages of the type (\ref{celav}) we introduce
spacelike unit vectors $e_1^\m$, $e_2^\m$, $e_3^\m$ that
form a tetrad together with $u^\m$, and define
$e^\m(\O) = \cos\varphi\, \sin\vartheta \, e^\m_1 + \ldots$ through
standard spherical coordinates $\O = (\varphi, \vartheta)$;
these quantities satisfy
%Using %the identities
\beq \< e^{\m_1}\, \cdots \, e^{\m_{2p+1}}\> =0,\quad 
    \< e^{\m} e^{\n}\> = \frac{1}{3}h^{\m\n},  \quad
    \< e^{\m} e^{\n}e^{\r} e^{\s}\> =
    \frac{1}{15}(h^{\m\n}h^{\r\s} + h^{\m\r}h^{\n\s} + h^{\m\s}h^{\n\r}). \eeql{eforms}
Note how Eq.\  (\ref{dk}) expresses the quantity $d(x,k)$,
which depends both on the spacetime coordinates $x^\m$ and the tangent space
coordinates $k^\m$, in terms of the tensor quantity $\D_{\m\n}$ (depending
\emph{only} on the $x^\m$) and the bilinear $k^\m k^\n$.
Therefore $\D^\mathrm{S}$, $\D^\mathrm{V}_\n$ and $\D^\mathrm{T}_{\m\n}$
do not depend on $e^\m$, % (because they do not depend on $k^\m$),
and one can directly apply (\ref{eforms}) to find
\beq \<d(x,k)\> = \D^\mathrm{S},\quad
   \<d(x,k)^2\> = (\D^\mathrm{S})^2
   + \frac{4}{3}h^{\m\n}\D_\m^\mathrm{V}\D_\n^\mathrm{V}
   + \frac{2}{15}\D_{\m\n}^\mathrm{T}h^{\n\r}\D_{\r\s}^\mathrm{T}h^{\s\m}.  \eeq

Returning to the specific form (\ref{Dmn}) of $\D_{\m\n}$, application of the
projection operators (in the `hatted' version)
gives $\hat\D^\mathrm{St} = 0$
(so that $\hat\D^\mathrm{S} = \hat\D^\mathrm{Ss}$) and
\bea 
\hat\D^\mathrm{S} &=& \frac{1}{3}\gh^{\m\n}\uh_{\m\sch\n}, \\
\hat\D_\m^\mathrm{V} &=& \2 \uh_{\m\sch\r}\uh^\r, \\
\hat\D_{\m\n}^\mathrm{T} &=&
\uh_{(\m\sch\n)} - \hat h_{\m\n} \hat\D^\mathrm{S} + \uh_\m \hat\D_\n^\mathrm{V}
   + \uh_\n \hat\D_\m^\mathrm{V},
%\frac{1}{3}h_{\m\n}h^{\r\s}\uh_{\r\sch\s} + \uh_{(\m} \uh_{\n)\sch\r}\uh^\r ,
\eea
i.e.\ these quantities correspond to the expansion, the acceleration and the
shear of the timelike vector field $\uh$ with respect to the metric $\gh$.

This has the following effects on the redshift.
In the integral in Eq.\ (\ref{redshfD}) we can write
$(\hat\D_{\m\n} / \hat u_\r)k^\m k^\n$ instead of $d(x,k) / (a\, u_\r)$.
Furthermore, since $(k^\m k^\n / k^\r) d \l$ is invariant under arbitrary
reparametrizations of the geodesic, we can replace it by
$(\tilde k^\m \tilde k^\n / \tilde k^\r) d \tilde \l$ with
$\tilde k^\m = \hat u^\m + \hat e^\m$
chosen such that $\hat u_\r \tilde k^\r = -1$ everywhere along the geodesic;
the factor $\hat\D_{\m\n}$ is unaffected because it depends only on $x$, not
on $k$.
Thus the argument of the exponential in Eq.\ (\ref{redshfD}) becomes
$\int_e^o \hat \D_{\m\n}\, \tilde k^\m \, \tilde k^\n d\tilde\l$.
Then, using the analog of Eq.\ (\ref{dDe}) for the metric $\hat g$,
%and the gauge conditions (\ref{gc}),
we get
\beq 1+z_{e\to o} = \frac{a(x_o)}{a(x_e)}
\exp\(\int_e^o (\hat \D^\mathrm{S} + 2 \hat \D^\mathrm{V}_\n \hat e^\n + \hat \D_{\m\n}^\mathrm{T}\,\hat e^\m \, \hat e^\n ) d\tilde\l\)
\eeql{redshgen}
%with $\hat \D_{\m\n}^\mathrm{T} = \uh_{(\m\sch\n)}$.
%
\del
This has the following effects on the redshift.
In the integral in Eq.\ (\ref{redshfD}) we can write
$(\hat\D_{\m\n} / \hat u_\r)k^\m k^\n$ instead of $d(x,k) / (a\, u_\r)$.
Furthermore, since $(k^\m k^\n / k^\r) d \l$ is invariant under arbitrary
reparametrizations of the geodesic, we can replace it by
$(\tilde k^\m \tilde k^\n / \tilde k^\r) d \tilde \l$ with
$\tilde k^\m$
chosen such that $\hat u_\r \tilde k^\r = -1$ everywhere along the geodesic.
Then we get
\beq 1+z_{e\to o} = \frac{a(x_o)}{a(x_e)}
\exp\(\int_e^o \hat \D_{\m\n}\, \tilde k^\m \, \tilde k^\n d\tilde\l\)\, 
\eeql{redshgen}
\enddel
for our preferred sources and observers whose worldlines have tangent
vectors $u^\m$.
If the actual emitter (`ae') and actual observer (`ao') have different
tangent vectors (but the same positions),
we must of course correct this via
\beq  1+z_{ae\to ao} =  (1+z_{ae\to e}) (1+z_{e\to o}) (1+z_{o\to ao}) ,\eeql{Doppler}
where $1+z_{ae\to e}$ and $1+z_{o\to ao}$ are just the standard
special-relativistic Doppler factors coming from the relative velocities
between the actual and preferred sources and observers, respectively.

\section{Gauge choice and metric}

The actual universe features deviations from homogeneity, so we do not
expect all components of $\D_{\m\n}$ to vanish.
Why can we nevertheless find a local frame in which the CMB has almost
exactly the same temperature in all directions?
We propose that this can be explained in the following manner.
Eqs.\ (\ref{redshgen}), (\ref{Doppler}) give the correct redshift for
arbitrary sources and observers and arbitrary functions $a(x)$ and vector
fields $u(x)$.
The result is of course independent of the choice of $a$ and $u$;
for most choices, several of the factors occurring
in Eqs.\ (\ref{redshgen}), (\ref{Doppler}) will get large or small,
and the computation
of the CMB redshift will involve cancellations between these factors.
If, however, we choose our setup such that $a(x)$ varies very little on the
last scattering surface, the relative velocities of the CMB sources are very
small, and the observer is the preferred one, then the only factor that
can still exhibit a strong direction dependence is
%strongly deviate from unity are %the direction-independent expression
%$a(x_o)/a(x_e)$ and
the exponential occurring in Eq.\ (\ref{redshgen}).
If we want to interpret the average of $a(x_o)/a(x_e)$, with the source
positions $x_e$ on the last scattering surface, as `the' redshift,
and every other factor as providing at most a further small fluctuation,
we need to ensure that the integral in Eq.\ (\ref{redshgen}) is small.
%is due to the fact that
We suggest to achieve this by choosing $a$ and $u$ in such a way that
\beq \D^\mathrm{S} = 0, \quad \D_\m^\mathrm{V} = 0, \eeql{gc} % to zero
which is an admissible \emph{gauge choice}.
Indeed, $\D^\mathrm{S}$ and $\D_\m^\mathrm{V}$ %these quantities
correspond to $1+3=4$ degrees of freedom, which is
just the number of quantities that can be fixed by a gauge.
%Then we get $d(x,k)=\D^\mathrm{T}_{\m\n}e^\m e^\n$ which has a vanishing
%expectation value.
\del
Then, using the analog of Eq.\ (\ref{dDe}) for the metric $\hat g$ and the
gauge conditions (\ref{gc}), and writing $\tilde k^\m = \hat u^\m + \hat e^\m$,
Eq.\ (\ref{redshgen}) becomes
\beq 1+z_{e\to o} = \frac{a(x_o)}{a(x_e)}
\exp\(\int_e^o \hat \D_{\m\n}^\mathrm{T}\,\hat e^\m \, \hat e^\n d\tilde\l\)\, 
\eeql{redshfDn}
with $\hat \D_{\m\n}^\mathrm{T} = \uh_{(\m\sch\n)}$.
\enddel
This choice reduces %the integrand in
the redshift formula (\ref{redshgen}) to
\beq 1+z_{e\to o} = \frac{a(x_o)}{a(x_e)}
\exp\(\int_e^o \hat \D_{\m\n}^\mathrm{T}\,\hat e^\m \, \hat e^\n d\tilde\l\).\, 
\eeql{redshfDn}
The tracelessness of $\hat \D_{\m\n}^\mathrm{T}$ together with statistical
isotropy ensures that the integrand
$\hat \D_{\m\n}^\mathrm{T}\,\hat e^\m \, \hat e^\n$
%this expression
has vanishing expectation value, and
in the next section we shall also see that it vanishes in linear
perturbation theory.
Thus it is not so surprising that the integral is small.

In terms of the original timelike field $u$, the effects of this choice
on the expansion
$\Theta = h^{\m\n}u_{\m ;\n}$, the acceleration $\dot u_\m = u_{\m;\r}u^ \r$,
the shear $\s_{\m\n} = u_{\m;\n}^\mathrm{PSTF}$
(the projected symmetric tracefree part of $u_{\m;\n}$, i.e.\ what remains
after symmetrizing, projecting with $h$ and removing the $h$-trace) and the
vorticity $\o_{\m\n} = h_\m{}^\r h_\n{}^\s u_{[\r;\s]}$ are easily found with the
help of Eq.~(\ref{cdu}):
\beq \dot u_\m = h_\m{}^\n\frac{a_{,\n}}{a}, \quad
   \Theta = 3 u^\r \frac{a_{,\r}}{a}, \quad
   \s_{\m\n} = a \uh_{(\m\sch\n)}, \quad
   \o_{\m\n} = a \uh_{[\m\sch\n]}.  \eeq
In words, expansion and acceleration correspond to the timelike and
spacelike components of $(\ln a)_{,\m}$, respectively;
shear and vorticity are multiples
of the corrresponding quantities in the conformally transformed frame.

\del
Let us return to the redshift experienced by
a photon emitted at $x_e$ and observed at $x_o$, with both source and observer
moving along worldlines whose tangent vectors are determined by $u$.
By virtue of Eq.\ (\ref{dkdef}) we get
\bea \ln[- a(x) (u\cdot k)(x)]_e^o
  &=& \int_e^o \frac{d(x,k)}{a(x) (u\cdot k)(x)} d\l \\
  &=& \int_e^o \frac{\D_{\m\n}\,k^\m\, k^\n}{a\, u_\r\,k^\r} d\l \\
  &=& \int_e^o \frac{\hat\D_{\m\n}\,\hat k^\m\, \hat k^\n}{\hat u_\r\,\hat k^\r}
      d\hat\l \\
  &=& -\int_e^o \hat \D_{\m\n}^\mathrm{T}\,\hat e^\m \, \hat e^\n d\hat\l \, ;\eea
in passing from the second line, where we still use the affine parametrization,
to the third, where $\hat k$ is chosen to satisfy $\hat u \cdot\hat k = -1$
we have used the fact that $(k^\m k^\n / k^\r) d \l$ is invariant under
reparametrizations of the geodesic;
for obtaining the last line we used Eqs.\ (\ref{dDe}) and (\ref{gc}).
This has the following effects on the redshift.
In the integral in Eq.\ (\ref{redshfD}) we can write
$(\hat\D_{\m\n} / \hat u_\r)k^\m k^\n$ instead of $d(x,k) / (a\, u_\r)$.
Furthermore, since $(k^\m k^\n / k^\r) d \l$ is invariant under arbitrary
reparametrizations of the geodesic, we can replace it by
$(\tilde k^\m \tilde k^\n / \tilde k^\r) d \tilde \l$ with
$\tilde k^\m = \hat u^\m + \hat e^\m$
chosen such that $\hat u_\r \tilde k^\r = -1$ everywhere along the geodesic.
\enddel

Let us now find explicit coordinates that implement our gauge
(\ref{gc}). %$\D^\mathrm{S}=0$, $\D_\m^\mathrm{V}=0$.
Choosing $\uh$ to be the vector with components $\uh^0=1$ and $\uh^i=0$,
we get $\gh_{00}=-1$,
$\uh^\m{}_{\sch\r} = \uh^\m{}_{,\r} + \Gh^\m{}_{\r\n}\uh^\n = \Gh^\m{}_{\r 0}$
and therefore
\beq \uh_{\m\sch\r} = \Gh_{\m\r 0}.  \eeql{uhGh}
Upon demanding $ 0 = 2\hat\D^\mathrm{V}_\m = \uh_{\m\sch\r} \uh^\r = %\Gh^\m{}_{00}$
%results in $0 =
\Gh_{\m 0 0} = \gh_{\m 0,0}$,  %\\   %$, and
%implies that $\gh$ is expansionless.
the metric takes the form $ds^2 = a^2\,d\hat s^2$ with
\beq d\hat s^2 = -(dx^0 - V_i\, dx^i)^2 + \g_{ij}dx^idx^j, \eeql{shat}
where $a$ and $\g_{ij}$ can depend on all coordinates $x^\m$ whereas $V_i$
depends only on the spatial coordinates $x^j$.
%One easily finds that t
The inverse metric $\gh^{\m\n}$ has the components
\beq \gh^{00} = -1 + V_i \g^{ij}V_j,\quad \gh^{0j} = \g^{jk}V_k,
   \quad \gh^{ij} = \g^{ij},
\eeq
where $\g^{ij}$ is defined by the  requirement $\g^{ij}\g_{jk} = \d^i_k$.
In matrix notation, the original metric and its inverse are given by
\beq g = a^2 \pmatrix{ -1 & V^T \cr V & \g - V V^T  } ,\qquad
g^{-1} = a^{-2}\pmatrix{ -1 + V^T \g^{-1} V & V^T \g^{-1}\cr \g^{-1} V & \g^{-1} } .
\eeql{matmet}
Finally,
$ 0 = 6\hat \D^\mathrm{S} = 2\gh^{\m\n}\uh_{\m\sch\n} = 2\gh^{\m\n}\Gh_{\m\n 0}
= \gh^{\m\n}(\gh_{\m\n,0} + \gh_{\m 0,\n} - \gh_{\n 0,\m}) = \gh^{\m\n}\gh_{\m\n,0}
%\label{ggdot}\eea
%Then Eq.~(\ref{ggdot}) becomes
= \gh^{ij}\g_{ij,0}
= \tr(\g^{-1}\g_{,0}) = (\tr\ln\g)_{,0} = (\ln\det\g)_{,0}$
implies $x^0$-independence of $\det \g$.

The conditions $V_{i,0} = 0$ and $(\det \g)_{,0} = 0$ do not completely
fix the form of our metric (\ref{shat}). %since
For example, they also hold in a transformed frame $\{\tilde x^\m\}$ with
\beq \tilde x^0 = x^0 + f(x^j), \quad \tilde x^i = \tilde x^i(x^j). \eeql{resgi}
We can use parts of this freedom to assign a single time coordinate to
%every point in the last scattering surface
%by a redefinition of the spatial coordinates in the initial surface,
%we can also achieve
the initial singularity and to set $\det \g = 1$.  %, \eeq
%which we assume from now on.
\del
Upon the standard split of $V$ into a gradient and a divergence free vector,
the former can be absorbed into a redefinition of $x^0$, so we may
assume that $V$ satisfies
\beq V_{i,0} = 0, \quad \d^{ij} V_{i,j} = 0. \eeq
The 1-form $\uh$ with components $\uh_\m = \gh_{\m\n}\uh^\n = \gh_{\m 0}$
is $\uh = -dx^0 + V_idx^i$.
Therefore, with (\ref{shat}), the projection operator
$\hat h_{\m\n} = \uh_\m \uh_\n + \gh_{\m\n}$ satisfies
$\hat h_{\m\n}dx^\m dx^\n = \g_{ij}dx^idx^j$.
In terms of this explicit frame we find that the non-vanishing components 
of shear and vorticity are given by
\beq    \s_{ij} = \frac{a}{2} \g_{ij,0}, \quad
   \o_{ij} = a V_{[i,j]}.  \eeq
\enddel

\section{Linear perturbation theory}

%To get a feeling for
We would now like to consider the consequences of our gauge choice (\ref{gc})
%$\D^\mathrm{S}=0$, $\D_\m^\mathrm{V}=0$,
in the context of linear perturbation theory \cite{Bardeen:1980kt}.
Our notation will be similar to that of Refs.\ \cite{Mukhanov:1990me,emm}
which we also recommend for further details.
A metric corresponding to a small perturbation of the conformally
flat case is given, before gauge fixing, by
\beq ds^2 = \ah^2(x^0) \{- (1+2\phi) (dx^0)^2 + 2 (B_{,i}-S_i)dx^i dx^0
+ [(1-2\psi) \d_{ij} + 2 E_{,ij} + 2 F_{(i,j)} + h_{ij}] dx^idx^j\};
\eeql{linmet}
here $\ah(x^0)$ represents the scale factor for the 
corresponding homogeneous case $(g_{\mathrm{h}})_{\m\n} = \ah^2 \eta_{\m\n}$;
$\phi$, $\psi$, $B$ and $E$ are scalars; $S_i$ and $F_i$ are transverse vectors
(i.e.\ they satisfy $\d^{ij}S_{i,j} = 0$ and $\d^{ij}F_{i,j} = 0$);
$h_{ij}$ is a symmetric traceless transverse tensor
($h_{ij} = h_{ji}$, $\d^{ij}h_{ij}=0$, $\d^{ik}h_{ij,k}=0$).
The gauge freedom $x^\m\to \tilde x^\m(x^\n)$ can be expressed at the
linearized level in terms of a transverse vector $\xi^i$ and scalars
$\xi^0$ and $\xi$; the corresponding transformations
\bea &\tilde \phi = \phi - \frac{\ah'}{\ah}\xi^0 -{\xi^0}_{,0},
\quad \tilde \psi = \psi + \frac{\ah'}{\ah}\xi^0,
\quad \tilde B = B + \xi^0 - \xi_{,0},
\quad \tilde E = E - \xi,&\\
& \tilde F_i = F_i -\xi_i,
\quad \tilde S_i = S_i + \xi_{i,0},
\quad \tilde h_{ij} = h_{ij} & \eea
can then be used to eliminate two of the scalars and
one of the transverse vectors.
The two most popular gauge choices are longitudinal gauge with $B=E=0$
(usually accompanied by neglecting vector and tensor modes), and synchronous
gauge, which manifests itself at the linearized level as
$\phi = B = 0$, $S_i=0$.

A well-known solution to the Einstein equations for irrotational dust
with $\L = 0$ (hence $\ah = \mathrm{const} \times (x^0)^2$),
which is believed to give a good description of the early matter dominated
era of our universe,
%given in terms
relies on a single time-independent function $\phi_\mathrm{N}$ which is just
the Newtonian potential.
In the longitudinal gauge this solution is given by
$\phi_\mathrm{long} = \psi_\mathrm{long} = \phi_\mathrm{N}$; it %This solution
can be transformed to the synchronous gauge via
$\xi^0 = x^0 \phi_\mathrm{N} /3$, $\xi = (x^0)^2 \phi_\mathrm{N} /6$,
resulting in 
$E_\mathrm{sync} = -(1/6) (x^0)^2 \phi_\mathrm{N}$,
$\psi_\mathrm{sync} = (5/3) \phi_\mathrm{N}$.
In the latter case, second derivatives of $\phi_\mathrm{N}$ occur
in the metric and tend to make the perturbations large for moderate $x^0$,
which is often used as an argument against employing the synchronous gauge
in situations other than the very early universe.

What about the gauge (\ref{gc}) %$\D^\mathrm{S}=0$, $\D_\m^\mathrm{V}=0$
and the corresponding metric (\ref{matmet})?
If we assume that we have used some of our residual gauge freedom to set
$\det \g =1$, then in the linearized version $\g_{ij} - \d_{ij}$ must be
traceless.
Writing $a = (1+\phi )\ah$, this implies $\d^{ij}E_{,ij}=3(\phi + \psi)$.
It turns out that without violating our gauge conditions we can set $B$ and
$S_i$ to zero, so that the metric becomes (up to quadratic and higher terms)
\beq ds^2 = \ah^2(x^0) (1+2\phi) \{- (dx^0)^2 
+ [\d_{ij} + 2 (E_{,ij} - \frac{1}{3}\d^{kl}E_{,kl}\d_{ij})
  + 2 F_{(i,j)} + h_{ij}] dx^idx^j\}.   \eeql{mymetp}
For the special solution considered above we
can get to this form by applying
a transformation with $\xi^0 = 0$, $\xi^i = 0$ and $\xi$
%chosen in such a way that it
satisfying $\xi_{,0} =0$ and $\d^{ij}\xi_{,ij} = - 6 \phi_\mathrm{N}$ to the
metric  in the longitudinal gauge.
This results in $\phi = \phi_\mathrm{N}$ and $E$ chosen such that
$\d^{ij}E_{,ij} = 6 \phi_\mathrm{N}$.
Thus we can interpret $E$ as a gravitational prepotential.
In particular, the expressions $E_{,ij}$ occurring in the metric should
be roughly of the same order of magnitude as $\phi_\mathrm{N}$.

It is instructive to apply our formalism to the metric (\ref{linmet})
that is not restricted by a gauge choice.
Considering the preferred observer to be the comoving one, we get
$a(x) = \ah(x^0)(1+\phi)$ and
\beq d\hat s^2 = - (dx^0)^2 + 2 (B_{,i}-S_i)dx^i dx^0
+ [(1-2\psi-2\phi) \d_{ij} + 2 E_{,ij} + 2 F_{(i,j)} + h_{ij}] dx^idx^j.
\eeql{linmethat}
Using Eq.\ (\ref{uhGh}), 
%If we ignore the vector and tensor modes,
we find
$\hat \D_{\m\n} = \uh_{(\m\sch\n)} = \Gh_{(\m\n) 0} = \2 \gh_{\m\n,0}$
for a general $\gh_{\m\n}$.
It is straightforward to compute and decompose this expression for
the metric (\ref{linmethat}), resulting in
\bea 
\hat\D^\mathrm{S} &=& -\psi_{,0}  -\phi_{,0} + \frac{1}{3}\d^{ij}E_{,ij0}, \\
\hat\D_i^\mathrm{V} &=& \2 (B_{,i} - S_i)_{,0}, \\
\hat\D_{ij}^\mathrm{T} &=& E_{,ij0} - \frac{1}{3}\d_{ij}\d^{kl}E_{,kl0}
   + F_{(i,j)0} + \2 h_{ij,0}.
\eea
We see again how the metric (\ref{matmet}) ensures the vanishing of
$\D^\mathrm{S}$ and $\D^\mathrm{V}$.
Expanding Eqs.\ (\ref{redshgen}), (\ref{Doppler}), with source and observer
velocities of $v^i_e$ and $v^i_o$, respectively, to the linear level,
results in
\beq  1+z_{ae\to ao} =  \frac{\ah(x_o)}{\ah(x_e)}\{1 + [\phi + v_i\hat e^i]_e^o
+ \int_e^o[-\psi_{,0} - \phi_{,0} + (B_{,i} - S_i)_{,0}\hat e^i
  + (E_{,ij0} + F_{(i,j)0} + \2 h_{ij,0})\hat e^i \hat e^j ] d\tilde \l \} .
\eeql{linredsh}
This expression is in full agreement with corresponding results in the
literature.
(To get, for example, Eq.\ (11) of Ref.\ \cite{Yoo:2009au}, one has to note
several different naming and
sign conventions including the directions of the unit vectors, and to
partially integrate the $(B_{,i} - S_i)$-term.)
As explained in detail in Ref.\ \cite{Yoo:2009au}, Eq.\ (\ref{linredsh})
contains all the standard contributions to the redshift, such as, for
example, the Sachs-Wolfe effect \cite{Sachs:1967er}.

For the dust solution considered above,
neither the longitudinal gauge nor the gauge advocated here lead to
corrections at the linearized level since the linearized fields are
$x^0$-independent in these gauges;
in contrast to this, the synchronous gauge features corrections 
because $E_\mathrm{sync} = -(1/6) (x^0)^2 \phi_\mathrm{N}$, in consistency
%expected since
with observations which show that the matter frame (the preferred frame
in the synchronous gauge) substantially differs
from the CMB frame.
While linear perturbation theory provides an excellent description of
the early universe, nonlinearities do play an important role in later eras,
and this is where we expect differences between the gauge (\ref{gc}) and
some nonlinearly consistent version of the longitudinal gauge such as the
Poisson gauge to manifest themselves.

In the simplified model mentioned above one could compute the source
velocities as the matter velocities $v_i = T_{0i}/T_{00} = G_{0i}/G_{00}$
from the components of the energy-momentum tensor and therefore from
the Einstein tensor, but this would neglect the different motions of
visible and dark matter.
A complete analysis of the CMB fluctuations would include an early,
perturbative part in which these and many more details are taken into
account; this would include the temperature variations, the actual source
velocities taking into account the incomplete alignment of dark and hadronic
matter, contributions of the radiation field to the energy-momentum
tensor, etc.
This can be done with the perturbative version (\ref{mymetp}) of the
metric (\ref{matmet}), or by transforming results obtained in any other gauge
to the present setup.
At a point in the history of the universe
where linear perturbation theory is still a good approximation
but radiation can already be neglected, one should then hand over to a
fully relativistic $\L$CDM simulation in the gauge (\ref{gc}).

Let us briefly summarize the results of this section.
The present formalism passes the consistency check of providing the
correct linearized redshift formula (\ref{linredsh}) in a general gauge.
Our metric is well behaved: in contrast to the synchronous gauge, the
linearized expressions do not exhibit a time dependence that would quickly
lead to troubles.
The integral occurring in the redshift formula (\ref{redshgen}),
which represents those deviations from the uniform case
that cannot be attributed to properties of the sources,
%is an average over a quantity that
vanishes at first order of perturbation theory in a simple matter-only
model, both in longitudinal gauge and in the gauge (\ref{gc}),
but only in the latter the first two contributions
$\hat\D^\mathrm{S}$ and $\hat\D^\mathrm{V}_\n\hat e^\n$ vanish at all orders.
%In view of this, the
The remaining quantity $\hat\D^\mathrm{T}_{\m\n} e^\m e^\n$ has an expectation
value of zero at all orders.
These facts make our formalism particularly useful for understanding why
we observe almost perfect isotropy of
the CMB %is not as surprising any more.
despite the existence of severe inhomogeneities in the non-linear era.

\section{Concluding remarks}

Observational cosmology relies not only on the redshift, but also on
other distance measures such as the angular diameter distance and the
luminosity distance.
These quantities can be computed via arguments based on fluxes.
For known redshift, one can use a comparison between the total number
of photons emitted per unit of time in a specific frequency range,
and the number of photons, in the appropriately transformed frequency range,
arriving in a given area at the observer's location.
Because of the non-acceleration and non-expansion of the vector field $\uh$
with resepct to $\gh$, the number of photons arriving per unit of $x^0$
(the time coordinate related to $\uh$) on a suitable hypothetic screen
enveloping the source must be identical with the number of photons emitted
during the corresponding $x^0$-interval of the same duration
(as measured with $\gh$).
Therefore, on average the photon count with respect to $\gh$ behaves like
the photon count in a static universe.
Upon proper rescalings of the time and area values with the corresponding
powers of $a$ one gets formulas for averaged fluxes that are identical in form
with those for a homogeneous universe, but with $\ah$ replaced by $a$.
Thus the overall expansion, as inferred from measured redshift-distance
relations, is given straightforwardly by the values of $a$ %=\ah\exp(\phi)$
at the sources and at our spacetime position.

There have been suggestions (for a small subset, see e.g.\
\cite{Schwarz:2002ba,Wiltshire:2007jk,Buchert:1999er,Rasanen:2006kp,Buchert:2007ik,Skarke:2015yza}) that
the perceived acceleration of the universe's expansion may not be due to
a cosmological constant or dark energy, but to some effect
stemming from the inhomogeneities of the actual universe.
%Claims of this type are
This possibility is rejected in papers such as \cite{Ishibashi:2005sj,Green:2014aga}, giving rise to
further rounds of controversy \cite{1505.07800,1506.06452}.
One of the main points of \cite{Ishibashi:2005sj,Green:2014aga} is an attack
on the choice of
synchronous gauge on which many attempts to explain the data without $\L$
are based; instead the use of the longitudinal gauge is advocated.
From %the point of view of
the present work it is clear that neither of these gauges is as directly
related to observations as the one presented here in Eq.~(\ref{gc}).

This makes a thorough investigation of the properties and consequences of
this gauge choice highly desirable.
Open questions include the following.
What residual gauge freedom is there beyond that indicated in (\ref{resgi})?
Is the possibility of setting $\gh_{0i} = V_i$ to zero %of the transformed metric
general or specific to linear perturbation theory?
What are the Einstein equations in linear and second order perturbation theory,
for collisionless dust and more generally?
Can we reproduce arguments along the lines of
\cite{Ishibashi:2005sj,Green:2014aga}?
What can we say beyond perturbation theory, either by analytic arguments or
numerically?

\noindent
{\it Acknowledgements:} 
It is a pleasure to thank Dominik Schwarz for discussions.

%\bye

%\small

%\enddel

\bye